\documentclass[preprint2]{aastex}

\shorttitle{EUV Emission from Abell\ 1795 and 2199}
\shortauthors{Bergh\"ofer \& Bowyer}

\begin{document}

\title{An Analysis of BeppoSAX LECS Observations of EUV Emission in Clusters
of Galaxies}

\author{Thomas W. Bergh\"ofer,\altaffilmark{1} Stuart
  Bowyer,\altaffilmark{2}}

\affil{Hamburger Sternwarte, Universit\"at Hamburg, Gojenbergsweg 
112, D-21029 Hamburg, Germany \and
Space Sciences Laboratory, University of California, Berkeley,
        CA 94720-7450, USA}

\begin{abstract}
Kaastra et al. (1999) have used the BeppoSAX LECS instrument to search for 
excess EUV emission in Abell\ 2199.  They claim that the results obtained 
confirm an independent report of an excess EUV emission in this cluster
(Lieu et al. 1999). Using an inflight derived procedure that is better suited 
to the analysis of extended sources and which avoids uncertainties related to 
ground-based calibrations for the overall detector sensitivity profile, we 
find no excess EUV emission in 
Abell\ 2199. We also used these procedures to search for an EUV excess
in Abell\ 1795, but no excess was found.
\end{abstract}

Subject headings: galaxies: clusters: individual (Abell 1795, Abell 2199) ---
techniques: image processing --- ultraviolet: general --- X-rays: general

\section{Introduction}
\label{intro}

The discovery of Extreme Ultraviolet (EUV) emission in clusters of galaxies 
with the Extreme Ultraviolet Explorer (EUVE) has provoked considerable 
controversy. While there is no doubt about the detection of the EUV 
emission in excess of that produced by the well-studied X-ray emitting cluster
gas in Virgo (Bergh\"ofer, Bowyer, \& Korpela 2000a) and in Coma
(Bowyer, Bergh\"ofer, \& Korpela 1999), many clusters do not exhibit an EUV
excess at least at current sensitivity levels. In a series of publications 
(Bowyer, Bergh\"ofer, \& Korpela, 1999; Bergh\"ofer, Bowyer, \& Korpela 
2000a,b; Bowyer, Korpela, \& Bergh\"ofer 2001) we have demonstrated that
the EUVE results are strongly affected by the variation of the telescope 
sensitivity over the field of view and upon the details of the subtraction of 
the EUV emission from the X-ray contribution. 

Kaastra et al. (1999) have analyzed BeppoSAX data obtained with the Low-Energy
Concentrator Spectrometer (LECS) to search for EUV emission in the Abell\ 2199 
cluster of galaxies.  Unfortunately, the telescope sensitivity profile used in
this work is likely to be incorrect.  Quoting from their work, Kaastra et 
al. (1999) state, "The vignetting correction for the LECS was derived from the
SAXDAS/LEMAT ray-trace code, assuming azimuthal symmetry around the appropriate
center. The correction for the support grid was also derived from that 
package." We note that ground based simulations of the large scale 
sensitivity of EUV and X-ray instrumentation are extraordinarily difficult to 
construct and are notorious for being an inappropriate representation of the 
true sensitivity functions.
In fact, a comparison of ground based calibration measurements with 
ray-tracing simulations for BeppoSAX clearly demonstrated a discrepancy by a 
factor of $\sim$1.5 at low energies (Parmar et al. 1997). 

In order to test the claim of an EUV excess in Abell\ 2199 by Kaastra et al. 
(1999) we have reanalysed BeppoSAX LECS observations of this cluster of 
galaxies. We also searched for excess EUV emission in Abell\ 1795 using 
archival
BeppoSAX data on this cluster. In Section\ \ref{data} we describe the data and
reduction applied to obtain the clusters' EUV emission. Section\ \ref{results}
provides the results of our investigations, and a summary is presented in 
Section\ \ref{discussion}.

\section{BeppoSAX LECS Data and Data Reduction}  
\label{data}

\begin{figure}
\plotone{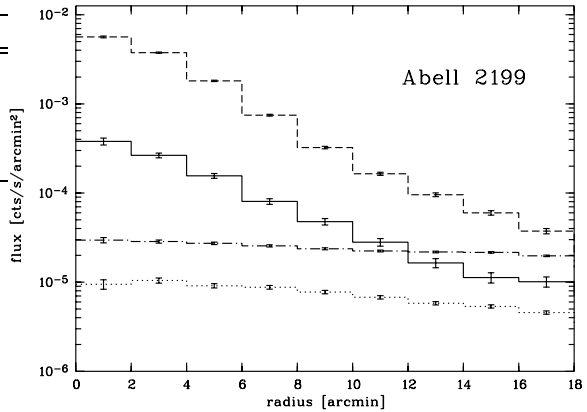}
\caption{Raw azimuthally averaged radial emission profiles of Abell\ 2199 in the
0.1--0.3\,keV (solid line) and 0.5--2.2\,keV (dashed line) energy bands of the
BeppoSAX LECS instrument. The dotted and dashed--dotted lines are the raw
background profiles obtained from the standard background files,
respectively, in the 0.1--0.3\,keV and 0.5--2.2\,keV band.}
\end{figure}

\begin{figure}
\plotone{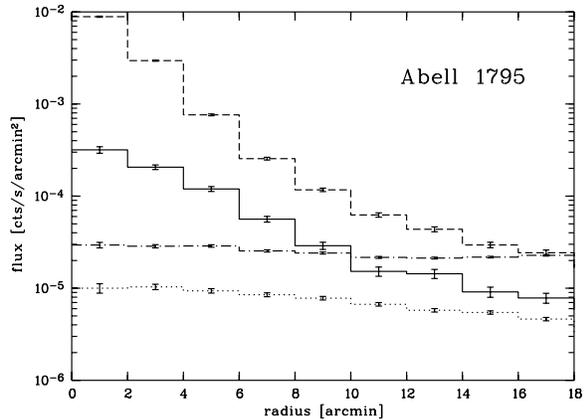}
\caption{Radial emission profiles of Abell\ 1795. The labeling of the lines
is the same as in Figure\ 1.}
\end{figure}

\begin{table}[tp]
  \caption{\label{obslog} BeppoSAX observation log.
}
  \begin{flushleft}
  \begin{tabular}{lcc}
    \hline
Target   & Exposure Date & Raw Exposure Time (s) \\
\hline\hline
Abell\ 1795   &  29 Dec 1996  & 3528 \\
             &  12 Aug 1997  & 7832 \\
             &  28 Jan 2000  & 38166 \\
Abell\ 2199   &  23 Apr 1997  & 26860 \\

\hline
  \end{tabular}
  \end{flushleft}
\end{table}

For our investigations we have employed BeppoSAX LECS archival data of the two
clusters Abell\ 2199 and Abell\ 1795. Table\ \ref{obslog} provides an observation
log of the analysed data sets.
The LECS instrument aboard of BeppoSAX is a gas scintillation proportional
counter especially designed to perform low-energy X-ray spectroscopy. A 
detailed description of this detector can be found in Parmar et al.
(1997). Here we summarize the performance of this detector relevant for
the study of diffuse sources in the low energy regime. The LECS has at least
some sensitivity in the energy range 0.1--10\,keV. Its circular field of view 
with a radius of about 18\arcmin\ is relatively small for cluster studies.
At 3 keV the on-axis angular resolution is roughly 3\arcmin\ (FWHM), which 
significantly drops to 9.7\arcmin\ (FWHM) at 0.28 keV. At these low energies
the resolution is dominated by the detector and is almost constant across the
field of view. At low energies the LECS effective area peaks near 0.2 keV. 
For our investigations we have selected all events in the detector channels 
10--30 ($\sim$0.1--0.3 keV)  and produced images of the clusters. The raw 
azimuthally averaged surface profiles of Abell\ 2199 and Abell\ 1795 and their 
statistical uncertainties are shown in Figures\ 1 and 2 (solid lines); in 
the case of Abell\ 1795 the plot is the combined profile of the three
distinct LECS observations of this cluster.

In order to determine the contribution of the X-ray emitting cluster gas
in the 0.1--0.3 keV band of the BeppoSAX LECS we also computed a radial
surface brightness profile at detector channels 50--200 ($\sim$0.5--2.2 keV). 
Events at
higher energies have been ignored due to the substantial drop in effective
area at energies E $>$ 2.5 keV and the increase in telescope vignetting at
higher energies, which might affect the clusters' X-ray profiles. 
In order to compensate for the substantially lower detector resolution at low 
energies, we convolved the source images in the 0.5--2.2 keV band with a
9.7\arcmin\ wide Gaussian. The derived 
radial emission profiles of the two clusters in the 0.5-2.2 keV band and its 
statistical errors are shown as dashed lines in Figures\ 1 and 2. 

For an appropriate background and sensitivity correction we obtained 
background files from the BeppoSAX data archive. In the case of the
LECS detector the standard background data consist of an assemblage of blank 
field observations with a total integration of 568562.8 seconds.
The dotted and dashed-dotted lines in Figures\ 1 and 2 provide the radial 
profiles obtained from the standard background data at the detector positions
of Abell\ 2199 and Abell\ 1795, respectively, in the 0.1-0.3 keV and 0.5--2.2 
keV band. 

Using procedures similar to the background subtraction method developed for 
the EUVE Deep Survey instrument (cf. Bowyer, Bergh\"ofer, \& Korpela 1999), we
adopted a two-parameter profile for the background. One profile is a flat 
background reflecting the time-dependent charged particle background. This has 
been determined from highly obscured regions at the outer most parts of the 
field of view. The count rates in units of $10^{-6}$cts\,s$^{-1}$\,arcmin$^2$ 
are 3.5, 5.1, and 6.3, respectively, for the background field and the science
exposures for Abell\ 2199 and Abell\ 1795.
The other profile reflecting all sensitivity changes over the 
detector field and telescope vignetting has been constructed from the standard 
background field. We subtracted the appropriate flat background and convolved
the data with a 9.7\arcmin\ wide Gaussian corresponding to the instrument
resolution; using the telescope point spread function from ground 
based calibration measurements instead of a Gaussian has no significant effect.
Scaling factors of 4.4 and 3.9 have been applied to the derived sensitivity 
profile to subtract this part of the background from the science exposures of 
Abell\ 2199 and Abell\ 1795. 
Note that this part of the background is dominated by scattered solar X-ray
radiation, which varies in time. Furthermore, the BeppoSAX observations used 
to construct the standard background data were taken at the beginning of the 
mission, whereas the cluster observations have been carried out closer to
solar maximum. This explains the relatively large scaling factors.
The given scaling factors and constant 
background rates result in the best representation of the background in the 
profiles. A large range of combinations of scaling factors and constant 
background rates were explored which were consistent with acceptable fits to 
the data, and even allowed for a possible contribution of an X-ray signal at 
larger radii. None of these variations changed our overall findings.

\section{Results}
\label{results}

\begin{figure}
\plotone{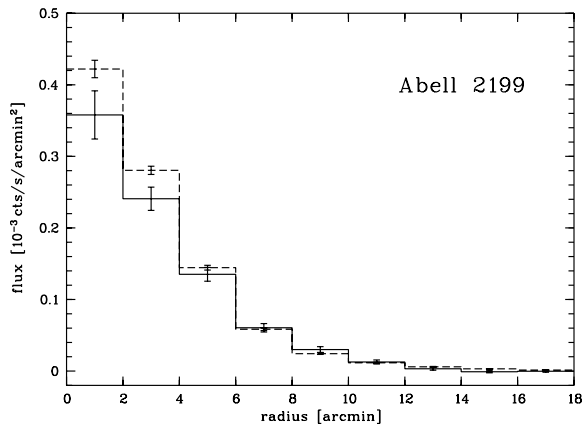}
\caption{Azimuthally averaged radial emission profiles of Abell\ 2199 in the
0.1--0.3\,keV (solid line) and 0.5--2.2\,keV (dashed line) band. The 
background has been subtracted from both profiles and the fluxes in the 
0.5--2.2\,keV were transformed into the 0.1--0.3\,keV band to reflect the
contribution of the X-ray emitting cluster gas in this energy band. There
is no excess EUV emission in this cluster.}
\end{figure}
In Figure\ 3 we show the EUV emission in the Abell\ 2199 cluster of galaxies 
obtained from BeppoSAX LECS data using the correct sensitivity profile 
(solid line). The EUV emission produced by the diffuse X-ray emitting cluster 
gas is shown by a dashed line. To transform the 0.5--2.2 keV band profile to 
the 0.1--0.3 keV band we used XSPEC and the detector response matrix of the 
LECS to simulate conversion factors for these two energy 
bands of the BeppoSAX LECS. Adopting the set of parameters and models for the 
X-ray emitting plasma and the absorption by the intervening interstellar 
medium as described in Bowyer, Bergh\"ofer, \& Korpela (1999) we found 
conversion factors of 10.6 and 9.9, respectively, for the 4.08 keV and 2.9 keV 
gas components.

Figure\ 3 demonstrates that the BeppoSAX LECS does not detect an EUV excess in
the cluster Abell\ 2199 when the data are analyzed correctly. We note that
there is a deficit of EUV emission in the central core region (R $<$ 4\arcmin)
of the cluster. However, with respect to the uncertainties in the cluster 
profiles, this is a small effect and may be a result of imperfect detector
calibration. On the other hand, such deficits has been observed in other 
clusters of galaxies and have been attributed in the past to the effects of  
cooling flows; in light of current results from XMM-Newton and Chandra, this 
effect must now be ascribed to some other cause (e.g., B\"ohringer et al. 2001,
Tamura et al. 2001).

The BeppoSAX LECS results on Abell\ 1795 are shown in Figure\ 4. The solid line
in this figure shows the background subtracted EUV emission profile of this 
cluster. The comparison with the expected contribution of the low energy tail 
of the X-ray emitting gas (dashed line in this figure) demonstrates that this 
cluster does not exhibit EUV excess emission at least at the BeppoSAX LECS 
level of sensitivity. Using appropriate model parameters for the X-ray 
emitting cluster gas and appropriate corrections for the interstellar 
absorption in the direction of Abell\ 1795 (see Bowyer, Berg\"ofer, \& Korpela 
1999), we applied conversion factors of 12.5 and 10.5, respectively, for the 
6.7 keV and 2.9 keV gas components. Again, the cluster center shows a small 
deficit in EUV radiation in the 2--4\arcmin\ radial bin when compared to
the expected cluster X-ray emission in the 0.1--0.3\,keV band. 
\begin{figure}
\plotone{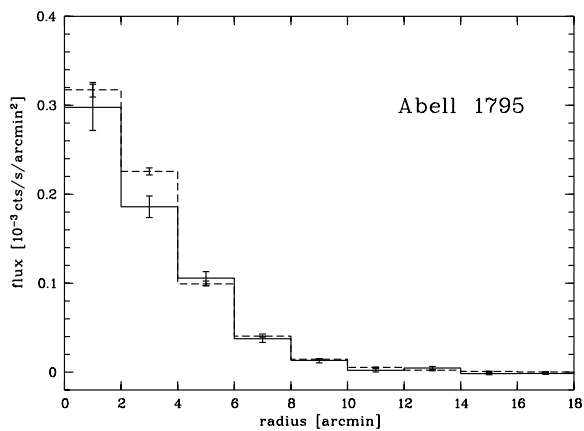}
\caption{Background subtracted radial emission profiles of Abell\ 1795 in the
0.1--0.3\,keV (solid line). The dashed line provides the contribution of
the X-ray emitting gas as can be obtained from the LECS 0.5--2.2\,keV band.
There is no excess EUV emission in this cluster.}
\end{figure}

\section{Discussion and Conclusion}
\label{discussion}

A search for excess EUV emission has been carried out in a substantial number 
of clusters observed with EUVE. Bowyer, Bergh\"ofer \& Korpela (1999) have 
shown that all but two of the reported detections were the product of the use 
of an incorrect detector sensitivity function. However, this conclusion has 
been questioned because Kaastra et al. (1999) claimed to have found an excess 
EUV emission in Abell 2199 using BeppoSAX LECS observations. These findings 
appeared to support the (incorrect) finding of an excess in this cluster using
EUVE data. However, this result was based upon ground-based estimates of the 
LECS detector sensitivity function. Using a procedure better suited to the 
analysis of extended sources that avoids the known uncertainties in the 
telescope sensitivity function, we show that there is no excess in Abell\ 2199.
We also searched for excess EUV emission in Abell\ 1795 using archival
BeppoSAX observations. No EUV excess was found.

The results obtained here for Abell\ 2199 and Abell\ 1795 are fully consistent 
with the results obtained on these clusters using EUVE data (Bowyer, 
Bergh\"ofer, \& Korpela 1999). The only clusters of galaxies that have been 
found to exhibit an excess EUV emission are the Virgo and Coma clusters 
(Bowyer, Korpela, \& Bergh\"ofer 2001).
It is possible that these are the only clusters that exhibit this effect, but 
it may be that both EUVE and BeppoSAX LECS are insufficiently sensitive to 
detect an EUV excess in other clusters of galaxies. 

Since the only clusters with a confirmed EUV excess are Virgo and Coma, it is 
useful to reconsider candidates for the underlying source of the EUV excess.
The original proposal was that this emission is thermal emission from a 
``warm'' (10$^6$ K) gas (Lieu et al. 1996a,b; Bowyer et al. 1996). Claims of 
``proof'' of this proposition have been advanced by Mittaz, Lieu, \& Lockman 
(1998), Lieu, Bonamente, \& Mittaz (1999a), Lieu et al. (1999b), Lieu, 
Bonamente, \& Mittaz (2000), and Bonamente, Lieu, \& Mittaz (2001a,b).
Lieu et al. (2000) misinterpreted small scale detector structures in the EUVE
data on Abell 2199 as cluster EUV emission absorbed by clumps of neutral 
hydrogen in the cluster. A difficulty with this interpretation which is 
independent of the data analysis problem is that the hydrogen required is 
"... $\sim$43 times more massive than the hot ICM in this region ... (and) ...
implies 3 times more missing baryons than expected''(op. cite).

The maintenance of a warm intracluster gas is quite difficult to understand 
since gas at this temperature is at the peak of its cooling curve and would 
cool in less than 0.5~Gyr, and on these grounds alone it was generally 
believed that a thermal source was untenable. Observational evidence relevant 
to this issue was obtained with the Hopkins Ultraviolet telescope (Dixon et 
al. 1996), and FUSE (Dixon et al. 2001a,b). No Far UV line emission from gas 
at 10$^6$ K was detected. More recently, observations of a large number of 
clusters with XMM have been carried out. Kaastra et al. (2001) found no gas at
T $<$ 1 keV in Sersic 159-03, Peterson et al. (2001) found no gas at 
T $<$ 2.7 keV in Abell 1835, and Tamura et al. (2001) found no gas at 
T $<$ 4 keV in Abell 1795. All other clusters observed with XMM showed no 
evidence of a cooler EUV emitting gas (Steve Kahn, private communication). 
The sum of this evidence seems overwhelming: a thermal mechanism for the EUV 
excess can be ruled out.

Since the underlying source mechanism is not thermal, it must be the product 
of some non-thermal process. Inverse Compton scattering of cosmic rays with 
the 2.7 K background was suggested early-on as a possible source mechanism 
(Hwang 1997, En{\ss}lin \& Biermann 1998). Sarazin \& Lieu (1998) suggested a 
model in which a population of cosmic rays produced several Gyr ago would have
degraded over time and would now be unobservable as radio synchrotron emission even at very low frequencies. This population would produce an EUV flux by
inverse Compton scattering. Sarazin \& Lieu derived the ratio between the 
azimuthally averaged total EUV emission and the azimuthally averaged soft 
X-ray flux predicted by their model; this ratio increases with increasing 
distance from the center of the cluster. Bergh\"ofer et al. (2000) derived 
this ratio for the Virgo cluster as a test of the Sarazin \& Lieu model. 
They found this ratio was flat with increasing distance from the center of the
cluster in contradiction to the prediction of the model.

Bowyer et al. (in progress) derived this ratio for the Coma cluster
using data on the cluster that had been analyzed correctly. They found
this ratio was flat with increasing distance from the center again
contradicting the predictions of the Sarazin \& Lieu model.

Despite the failure of the Sarazin \& Lieu model, the inverse Compton mechanism
remains as the only candidate for the source mechanism for the EUV excess. 
However, a new difficulty for this hypothesis has recently appeared. Virtually
all models invoking the inverse Compton mechanism require the intracluster 
magnetic field to be $<<$1$\mu$G. However, recent results show that cluster 
magnetic fields are quite large. Clarke, Kronberg, \& B\"ohringer (2001) 
studied 16 clusters with very high spatial resolution and have shown that all 
of these clusters have B fields of 4 to 7$\mu$G. Unless this result is
somehow incorrect, the vast majority of models proposed for the production of 
the EUV excess are incorrect. The only exceptions to the low field models 
(En{\ss}lin, Lieu, \& Biermann 1999; Atoyan \& V\"olk 2000) are unlikely to be
appropriate (Ming, Hwang, \& Bowyer 2001).

It is not clear whether this is a fundamental obstacle for the inverse
Compton scattering hypothesis or if it is simply a failure of existing
models. Irrespective of how widespread the occurrence of EUV excess in 
clusters of galaxies may be, the underlying source mechanism for this emission
remains a mystery.

\acknowledgments
This work was supported in part by a University of California grant. TWB was 
supported in part by a Feodor-Lynen Fellowship of the 
Alexander-von-Humboldt-Stiftung.

\end{document}